\documentclass[aps,prl,preprint,superscriptaddress]{revtex4-2}
\usepackage{graphicx}
\graphicspath{{Figures/}}
\usepackage{layouts}
\usepackage{siunitx}
\usepackage{xcolor}
\usepackage{textgreek}
\usepackage{amsmath}
\usepackage{tikz}
\newcommand*\circled[1]{\tikz[baseline=(char.base)]{
		\node[shape=circle,draw,inner sep=2pt] (char) {#1};}}

\begin{document}

% Use the \preprint command to place your local institutional report
% number in the upper righthand corner of the title page in preprint mode.
% Multiple \preprint commands are allowed.
% Use the 'preprintnumbers' class option to override journal defaults
% to display numbers if necessary
%\preprint{}

%Title of paper
\title{Oscillatory droplet dissolution from competing Marangoni and gravitational flows}

% repeat the \author .. \affiliation  etc. as needed
% \email, \thanks, \homepage, \altaffiliation all apply to the current
% author. Explanatory text should go in the []'s, actual e-mail
% address or url should go in the {}'s for \email and \homepage.
% Please use the appropriate macro foreach each type of information

% \affiliation command applies to all authors since the last
% \affiliation command. The \affiliation command should follow the
% other information
% \affiliation can be followed by \email, \homepage, \thanks as well.
\author{ Ricardo Arturo Lopez de la Cruz}
\affiliation{Physics of Fluids Group, Max-Planck-Center Twente for Complex Fluid Dynamics, Mesa+ Institute, and J. M. Burgers Centre for Fluid Dynamics, Faculty of Science and Technology, University of Twente, P.O. Box 217, 7500 AE Enschede, The Netherlands}
%\email[]{Your e-mail address}
%\homepage[]{Your web page}
%\thanks{}
%\altaffiliation{}
\author{Christian Diddens}
\affiliation{Physics of Fluids Group, Max-Planck-Center Twente for Complex Fluid Dynamics, Mesa+ Institute, and J. M. Burgers Centre for Fluid Dynamics, Faculty of Science and Technology, University of Twente, P.O. Box 217, 7500 AE Enschede, The Netherlands}
\author{Xuehua Zhang}
\affiliation{Department of Chemical and Materials Engineering, University of Alberta, Edmonton, Alberta, T6G 1H9, Canada}
\affiliation{Physics of Fluids Group, Max-Planck-Center Twente for Complex Fluid Dynamics, Mesa+ Institute, and J. M. Burgers Centre for Fluid Dynamics, Faculty of Science and Technology, University of Twente, P.O. Box 217, 7500 AE Enschede, The Netherlands}
\author{Detlef Lohse}
\affiliation{Physics of Fluids Group, Max-Planck-Center Twente for Complex Fluid Dynamics, Mesa+ Institute, and J. M. Burgers Centre for Fluid Dynamics, Faculty of Science and Technology, University of Twente, P.O. Box 217, 7500 AE Enschede, The Netherlands}
\affiliation{Max Planck Institute for Dynamics and Self-Organization, Am Fa\ss berg 17, 37077 G\"{o}ttingen, Germany}

%Collaboration name if desired (requires use of superscriptaddress
%option in \documentclass). \noaffiliation is required (may also be
%used with the \author command).
%\collaboration can be followed by \email, \homepage, \thanks as well.
%\collaboration{}
%\noaffiliation

\date{\today}

\begin{abstract}
% insert abstract here

The dissolution or growth of a droplet in a host liquid is an important part for processes like chemical extraction, chromatography or emulsification. In this work we look at the dissolution of a pair of  vertically aligned droplets immersed in water, both experimentally and with numerical simulations. The liquids used for the droplets are long chain alcohols with a low but finite solubility in water and a significantly lower density than that of the host liquid. Therefore, a solutal plume is formed above of the bottom droplet and natural convection dominates the dissolution process. We monitor the volume of the droplets and the velocity field around them over time. When the liquids of the two droplets are the same, our previously found scaling laws for the Sherwood and Reynolds numbers as functions of the Rayleigh number (Dietrich \textit{et al.}, 2016, \textit{J. Fluid  Mech.}) can be applied to the lower droplet. However, remarkably, when the liquid of the top droplet is different than that of the bottom droplet the volume as function of time becomes non-monotonic, and an oscillatory Marangoni flow at the top droplet is observed. We identify the competition between solutal Marangoni flow and density driven convection as the origin of the oscillation, and numerically model the process.

\end{abstract}

% insert suggested keywords - APS authors don't need to do this
%\keywords{}

%\maketitle must follow title, authors, abstract, and keywords
\maketitle

% body of paper here - Use proper section commands
% References should be done using the \cite, \ref, and \label commands

\section{Introduction}

Droplet dissolution, and more generally mass transfer between droplets and the surrounding liquid, are important processes for varying technological applications. For instance, any process industry in which emulsions are part of the processes will be concerned with droplet dissolution. One example is within contactor columns where mass transfer can affect the coalescence of the droplets of the emulsion \cite{groothuis1960influence}. As another example, many foods and cosmetics are in the form of emulsions; thus stabilizing these emulsions is important for the shelf life of the products \cite{zhang2020gelatins, tchakalova2014}. Yet another example is in the pharmaceutical industry where emulsions are used as a way to create drug delivery systems \cite{lepeltier2014nanoprecipitation}. In environmental technology, the dissolution of CO$_2$ droplets deep in the ocean has been seen as a way to reduce the concentration of this gas in the atmosphere \cite{brewer2002experimental, hirai1997dissolution}. Finally, for chromatography the mass transfer of a solute into a small droplet by liquid-liquid microextraction has been considered as a way to obtain high concentrations of the solutes needed \cite{jain2011recent, rezaee2006determination, rezaee2010evolution, lohse2016towards}. 

The examples listed above show how important droplets are for the process industry. Therefore, many efforts have been undertaken to better understand the diffusive dynamics of droplets, see an extensive discussion thereof in our reviews \cite{lohse2015surface, Lohse_2020}. On the theoretical side the classical diffusion based model for bubble dissolution of Epstein and Plesset \cite{epstein1950stability} has been extended to the dissolution of spherical pure \cite{duncan2006microdroplet} and binary \cite{su2013mass} droplets. Moreover, very recently, a more accurate binary droplet dissolution model was developed by Chu and Prosperetti \cite{chu2016dissolution} for the diffusion dominated regime. Instead of approximating the equilibrium conditions of the dissolving entities at the interface, it explicitly makes use of the equality of chemical potentials and considers non-ideal solutions, so that it could also be applied to translating droplets. Simultaneously, various experimental and numerical approaches for the dissolution of pure \cite{dietrich2015stick, dietrich2016role, zhang2021dissolution, Kovalchuk2006marangoni, su2010effect, sato2000direct, lappa2004mixed, chong2020convection, poesio2009dissolution, xie2019effect, basu2021dissolution}, binary \cite{dietrich2017segregation, duncan2006microdroplet}, and multicomponent droplets \cite{Tan_2019, Lohse_2020, maheshwari2017molecular} have been developed.

In the absence of an external flow, the dissolution of a droplet is purely determined by diffusion. The Epstein-Plesset like model has been shown to be successful in capturing experimental observations for such pure diffusive situations \cite{duncan2006microdroplet}. However, once either forced or natural convection is present, models based on diffusion only obviously fail to properly predict the dissolution time of droplets \cite{dietrich2016role, dietrich2017segregation, chong2020convection}. Dietrich \textit{et al.} \cite{dietrich2016role} showed that when natural convection is present, the dissolution time of a pentanol droplet was reduced to 3 \si{\hour} instead of the 11 \si{\hour} as predicted by the Epstein-Plesset model. This was because of the emergence of a low density plume of dissolved liquid on top of the droplet that accelerated the mass transport from the droplet into the host liquid. This process could quantitatively be modeled \cite{dietrich2016role} and numerically be simulated \cite{chong2020convection}. 

When two or more droplets sit next to each other, their mutual interaction can affect the dissolution time \cite{laghezza2016collective,bao2018flow, chong2020convection}. Chong \textit{et al.} \cite{chong2020convection} showed that when multiple droplets collectively dissolve in the purely diffusive regime, their dissolution time is increased because of shielding, as the dissolution of neighboring droplets reduces the concentration gradient experienced by the droplets. Moreover, if the droplets are arranged in a rectangular $3\times3$ lattice, due to shielding the central droplet takes longer to dissolve than those at the edges of the array. On the contrary, in the convective regime, the plumes created by each droplet can merge into one big centered plume, provided they are close enough, remarkably causing the dissolution time to be shorter than for a single droplet in the diffusive regime. But still the central droplet dissolves slower than those at the edges. Bao \textit{et al.} \cite{bao2018flow} showed that if instead of natural convection there is an externally imposed flow, the dissolution of a lattice of droplets also displays the shielding effect and the droplets at the center of the lattice are the last to dissolve, while those which are upstream dissolve first. Additionally, it was shown that not unsurprisingly higher flow velocities and a larger spacing between the droplets resulted in shorter dissolution times.

Alongside the phenomena described so far, the presence of non-uniform temperature \cite{young1959motion, bassano2003numerical, lappa2004mixed,  lappa2004higher, lappa2006oscillatory} and of concentration fields \cite{Li_2019_Bouncing, li2021marangoni, schwarzenberger2015relaxation} can further complicate the dynamics of a dissolving droplet by giving rise to thermal and solutal Marangoni flows. If the droplet is not fixed in place, the forces caused by the Marangoni flow can propel the droplet against its own weight. Depending on the properties of the system, this can result in the droplets being suspended at a higher location than where a buoyancy balance would suggest or even more surprisingly make the droplet repeatedly jump \cite{Li_2019_Bouncing}. In other cases, even without externally imposing a non-uniform field, the dissolution of the droplet can create  gradients in concentration around the droplet that result in self-propulsion \cite{michelin2013spontaneous, maass_2016, izri2014self, moerman2017solute, chen2021instabilities, poesio2009dissolution}.

In the present work we study the simultaneous dissolution of two droplets in the natural convection regime. Different to the case of Chong \textit{et al.} \cite{chen2021instabilities}, one of the droplets is placed above the other, such that the plume caused by the lower droplet rises towards the upper one. In this way we further study the effects of convection on the dissolution of droplets, but also the effect of having solute already dissolved in the host liquid. Moreover, by changing the liquid of the upper droplet, gradients in the interfacial tension can arise, leading to some Marangoni flow that, as we will see, can become oscillatory.  

The paper is organized as follows: in the next section we describe our experimental methods and the numerical setup with its corresponding model equations. Afterwards, we present the experimental results of the dissolution of single droplets and pairs of vertically aligned droplets. For the latter case, we study the effect of changing the liquid of the upper droplet. For some cases there is an oscillatory flow which we study in more detail numerically. We then discuss the mechanisms behind the observed oscillatory flow. The paper closes with the conclusion and outlook section.

\begin{figure}
	\centering
	\includegraphics[width=\textwidth]{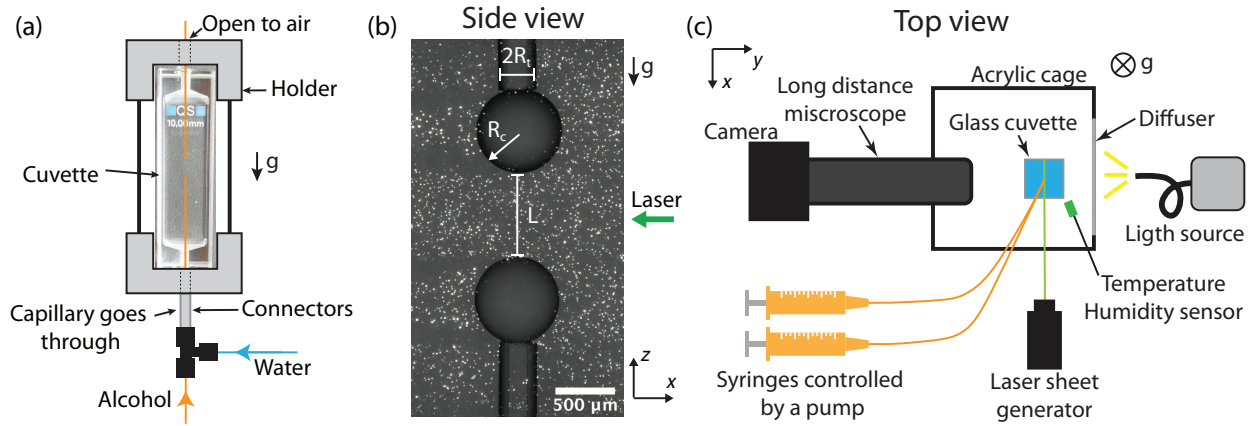}
	\caption{(a) Sketch of the cuvette and the capillaries connected to it. The bottom capillary passes through a T junction and wide-enough connectors into the cuvette. From the third leg of the T junction water is supplied into the cuvette. The water flows around the capillary inside the wide connectors. (b) Typical example of two droplets just after infusion with tracer particles. The definitions of three length scales are shown. The laser sheet comes from the right. (c) Sketch of the experimental setup from a top view. The glass cuvette was illuminated from the back (from the right in the image) with a white light source. A diffuser was used to have an approximately homogenous illumination. A laser sheet illuminated the central plane perpendicular to the direction of white light. A long distance microscope was used to observe the droplets. Each capillary was connected to a syringe mounted to a syringe pump. A cage of acrylic was built around the cuvette to reduce air currents and heat exchange with the surroundings. The top of the cage was open. }
	\label{fig:Setup}
\end{figure}

\section{Experimental methods \label{SecCh3:ExpMethods}}

\subsection{Experimental setup}

A cuvette (flow-trough cell, Suprasil quartz cuvette, Hellma) was used as a container, see Figure \ref{fig:Setup}(a). The cuvette had a square base and rectangular sides. The inner dimensions of the square were 10 \si{\milli\meter} $\times$ 10 \si{\milli\meter}. The total volume was 3300 \si{\micro\liter}. Both at the bottom and top there were inlets through which two fused silica capillaries were inserted (Polymicro Technologies, nominal inner diameter $150$ \si{\micro\meter}, nominal outer diameter $360$ \si{\micro\meter}). The lower capillary passed trough the straight section of a T junction. A third capillary was connected to the perpendicular arm of the T junction and was used to fill the cuvette with water. The upper inlet of the cuvette was not sealed so that we could move the top capillary for alignment. In Figure \ref{fig:Setup}(a) we include a sketch of the capillaries and connectors. The tips of the capillaries were slowly sanded to obtain an approximately flat end. The outside of the capillaries was covered with a polymer film, this film was removed close to the tips where the droplets were infused resulting in an outer diameter of 320 \si{\micro\meter}.

The capillaries were connected to two glass syringes (Hamilton, 100 \si{\micro\liter}, 1710 TLL,  no stop) controlled by a syringe pump (Harvard Instruments, PHD 2000). With the pump, droplets of 0.2 \si{\micro\liter} were infused and left to dissolve in a bath of water (see Figure \ref{fig:Setup}(b) for an example of the two droplets). For some experiments either the top or bottom capillary were removed in order to observe the dissolution of a single droplet. During the experiments, a small amount of evaporation took place at the top, but this was normally in the order of \mbox{3 \si{\micro\liter}} to 6 \si{\micro\liter} ($\sim$ 0.1\% the total volume). 

The cuvette was held in front of a long distance microscope (Navitar 12x) and illuminated from behind with light that passed trough a diffuser as depicted in Figure \ref{fig:Setup}(c). Videos were recorded with a charge-coupled device (CCD) camera (MD061MU-SY, Ximea). To measure the velocity field in the bath, a laser sheet was created using a La Vision system and a 532 \si{\nano\meter} laser of 0.1\si{W}. The power was set between 53 \% and 57 \%, with the majority at 55\%. The width of the laser sheet was 310 \si{\micro\meter} $\pm$ 70 \si{\micro\meter} during one set of experiments and 390 \si{\micro\meter} $\pm$ 10 \si{\micro\meter} during a second set of experiments, measured using thermal paper and a calibrated picture of the mark. Fluorescent particles (Fluoro Max, red fluorescent polymer microspheres 6 \si{\micro\meter}) were used to seed the flow with a concentration of $3\times 10^{-2}$ wt \%. The Stokes number for a pentanol-in-water solution and the employed tracer particles was of the order of $St\sim 10^{-6}$ to $10^{-4}$, calculated as in Ref. \cite{Li_2019_Bouncing}, taking as length scale the initial radius of the droplet and as characteristic velocity the Marangoni flow velocity at the interface of the upper droplet (as obtained from simulations as will be shown below) or the velocity of the plume, respectively. Since the $St$ number was much smaller than 1, the particles can be considered as faithful tracers. The particle image velocimetry (PIV) analysis was carried out with the software PIVlab in Matlab \cite{thielicke2014pivlab, Thielicke2019, Thielicke2014PhD}.

We noticed that when the experimentalist was standing close to the cuvette, a small flow inside the cuvette started due to thermal disturbances from the surrounding. Therefore, a cage of acrylic panels was built around the cuvette to reduce such thermal effect (see Figure \ref{fig:Setup}(c)) and the experimentalist stood away from the setup once an experiment started. In all our experiments the temperature and humidity around the cuvette were monitored with a sensor (HIH6130, Honeywell) placed a few centimeters from the cuvette as shown in the sketch in Figure \ref{fig:Setup}(c).

\subsection{Liquids used}
Liquids: The bath consisted of Milli-Q water (produced by a Reference A+ system (Merck Millipore) at 25\si{\degreeCelsius} and 18.2 M\textOmega cm). 1-Pentanol (ACS reagent $\leq $ 99\%), 1-Hexanol (reagent grade, 98\%), 1-Heptanol (98\%), and 1-Octanol (anhydrous $\leq $ 99\%) were bought from Sigma-Aldrich and used as received. The properties of the liquids are listed in Table \ref{tabCh3:PropertiesOfLiquids}.

\subsection{Experiment preparation}
The capilllaries were plasma-cleaned for 10 min before a set of experiments. After one experiment, about 0.5 \si{\milli\liter} of water was pumped to reduce the amount of pentanol dissolved. The water used in experiments was left at least 1 hour in the laboratory before use. When a different liquid was used the cuvette was sonicated in a mixture of water and acetone for 10 min, then dried. The connectors were sonicated in water for 10 min and then also dried.

\begin{table}
	\begin{center}
		\begin{tabular}{p{2.9cm}cccccccc}
			\hline
			Liquid & $\rho$ [\si[per-mode = fraction]{\kilo\gram\per\cubic\meter}] & $D$ [\si[per-mode = fraction]{\milli\meter\squared\per\second}] & $c_\mathrm{sat}$ [wt \%]& $\Delta \rho$ [\si[per-mode = fraction]{\kilo\gram\per\cubic\meter}] & $\mu$ [\si{\milli\pascal\second}] & $\gamma$ [\si[per-mode = fraction]{\milli \newton \per \meter}] & $\partial \gamma / \partial c$ [\si[per-mode = fraction]{\milli\newton \per \meter}] & Bo \\ [2pt]
			\hline 
			1-pentanol& $811^a$ & $888^b$ & $2.2^c$ & $3.42^a$ & $3.52^d$ & $4.4^e$ & -- & 0.05 \\
			1-hexanol & $814^a$ & $830^b$ & $0.6^c$ & $0.92^a$ & $4.40^d$& $6.8^e$ & -11.7& 0.03\\
			1-heptanol & $822^a$ & $800^b$ & $0.17^c$ & $0.29^a$ & $6.00^d$& $7.7^e$ & -20.4& 0.03\\
			1-octanol & $827^a$ & $780^a$ & $0.049^c$ & $0.07^a$& $7.60^d$ & $8.52^e$ & -30.9& \\
			water  & $997^f$& -- & -- & -- & $0.89^f$& -- & --& \\
			water saturated with pentanol& $994^f$ & -- & -- & -- & $0.98^f$& & & \\
			\hline& & 
		\end{tabular}
		\caption{Properties of the liquids (at 25 \si{\celsius}) used in experiments. $\rho$ is the density, $D$ is the diffusion coefficient, $c_\mathrm{sat}$ is the saturation mass fraction of the alcohols in water, $\Delta \rho$ is the difference in density between pure water and water saturated with the alcohol, $\mu$ is the dynamic viscosity, $\gamma$ is the interfacial tension with water, and $\partial \gamma / \partial c$ is the change in interfacial tension with mass fraction of pentanol as measured by the pendant droplet method (see the SI). The Bond number $Bo =  gR_{e,0}^2\Delta \rho / \gamma$ is obtained for an initial equivalent radius $R_{e,0} = 363$ \si{\micro\meter} and the density difference between pure water and the corresponding alcohol.	$^a$Obtained from Ref. \cite{dietrich2016role}. $^b$Obtained from Ref. \cite{hao1996binary}. $^c$Obtained from Ref. \cite{kinoshita1958solubility}. $^d$Obtained from Ref. \cite{al2004densities}. $^e$Obtained from Ref. \cite{demond1993estimation}. $^f$Obtained or extrapolated from Ref. \cite{pai1998viscosity}.}
		\label{tabCh3:PropertiesOfLiquids}
	\end{center}
\end{table}

\subsection{Numerical setup \label{subsecCh3:NumSetup}}

\begin{figure}
	\centering
	\includegraphics[width=0.9\textwidth]{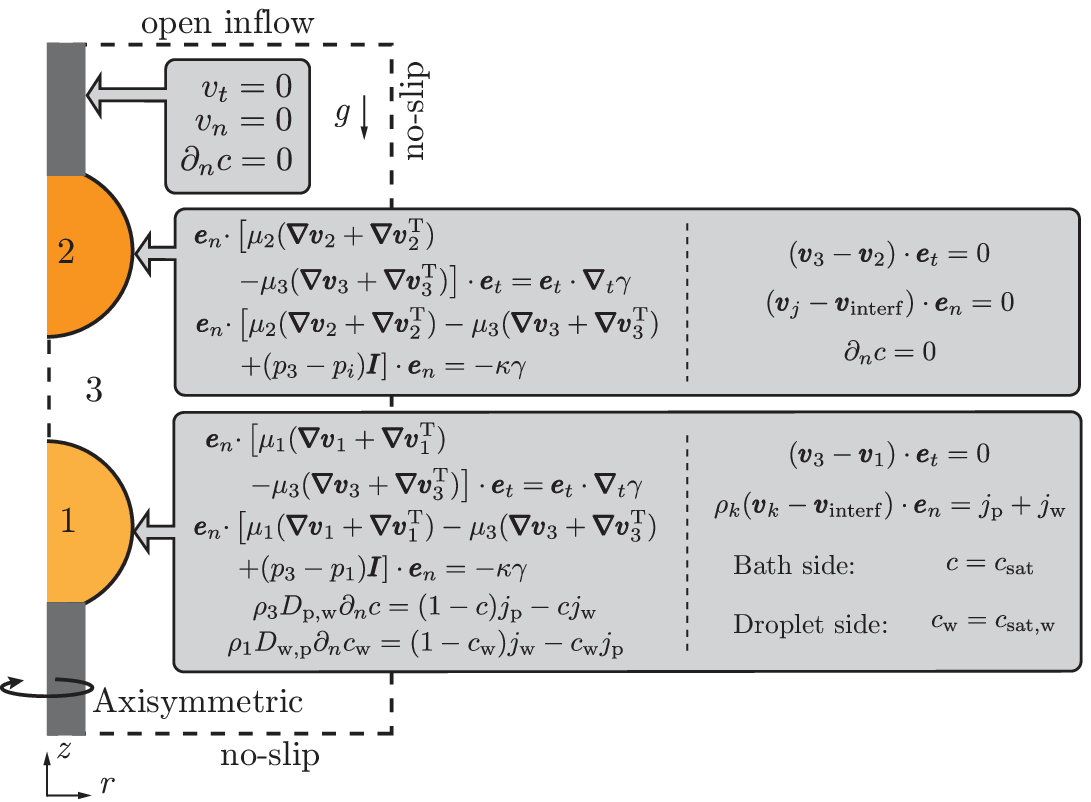}
	\caption{Schematic of the numerical setup with the boundary conditions. The subscripts $j = 2,3$ and $k = 1,3$. }
	\label{fig:SimBC}
\end{figure}

We performed numerical simulations for the case of a 1-pentanol droplet at the bottom and a 1-octanol droplet at the top. We used a finite element method implemented on the basis of the finite element package \textsc{oomph-lib} \cite{Heil2006}. The liquids inside the droplets and the surrounding bath are discretised using triangular Taylor-Hood elements for the degrees of freedom of the velocity and pressure space, and a linear shape function for the composition in the three domains. The general numerical technique is described in detail in Ref. \cite{Diddens2017c}, however, in that work the droplet was assumed to be a spherical cap while here the full interface dynamics are solved, thus the interface can deform and  deviate from the shape of a spherical cap. Comparison with various experimental cases for benchmarking of the code can be found in \cite{Diddens2017, Li2019, Li_2019_Bouncing}. 

The cylindrical symmetry of the system allows us to use an axisymmetric simulation as shown in Figure \ref{fig:SimBC}. The chosen domain is cylindrical, with 8 \si{\milli\meter} in height and 3 \si{\milli\meter} in diameter. Two solid cylinders are placed at the top and bottom along the axis of symmetry. As opposed to the experiments, the cylinders are not capillaries, thus the droplets are not connected to a reservoir of 1-pentanol. The base of the droplets is pinned to the cylinders. No-slip boundary conditions, a zero normal velocity and no mass transfer were considered at the edges of the cylinders as depicted in Figure \ref{fig:SimBC}.

Apart from the cylinders, the domain was divided into three regions: 1 the bottom droplet, 2 the top droplet, and 3 the bath. In the three regions we considered the continuity equation for incompressible fluids and the Navier-Stokes with the Boussinesq approximation,
\begin{gather}
	\pmb{\nabla\cdot v}=0,\\
	\rho_\alpha\left(\partial_t \pmb{v}_\alpha + \pmb{v}_\alpha \cdot\pmb{\nabla v}_\alpha \right) = -\pmb{\nabla}p_\alpha + \pmb{\nabla}\cdot[\mu_\alpha (\pmb{\nabla v}_\alpha+\pmb{\nabla v}^T_\alpha)] -\rho_\alpha g \pmb{e_z},
\end{gather}
where $v_\alpha$ and $p_\alpha$ are respectively the velocity and pressure in each region $\alpha = 1, 2, 3$ and $g$ the acceleration of gravity. The density and viscosity are functions of the mass fraction of 1-pentanol $c$, i.e. $\rho_\alpha = \rho_\alpha(c)$, and $\mu_\alpha=\mu_\alpha(c)$. The superscript $T$ indicates the transpose. The reason behind the use of the Boussinesq approximation for the continuity equation lies in the numerics. While the complete equation considers density variation accurately, the partial masses are not perfectly conserved within the numerical scheme. However, when considering a divergence free fluid, this is not an issue, so we have decided to choose the Boussinesq approximation.

Inside the bottom droplet and the bath the advection diffusion equation is solved,
\begin{equation}
	\rho_\alpha(\partial_t c + \pmb{v} \cdot \pmb{\nabla} c)=\pmb{\nabla}\cdot(\rho_\alpha D \pmb{\nabla}c),
\end{equation}
where $D$ is the diffusion coefficient of either pentanol in water (in the bath) or of water in pentanol (inside the lower droplet). Solving for either the mass fraction of 1-pentanol $c$ (in the bath) or of water $c_\mathrm{w}$ (inside the droplet) is enough as they are coupled by the relationship $c_\mathrm{w} + c = 1$. As a simplification, the top droplet is considered insoluble in water, thus there is no need to solve for the concentration inside the top droplet, and there is no third component going into the bath. The implications of this simplification will be addressed in the discussion section below.

The boundary conditions for the tangential and normal stresses at the interfaces of both droplets are given by
\begin{gather}
	\pmb{e}_n\cdot  \left[\mu_i(\pmb{\nabla v}_i+\pmb{\nabla v}_i^\mathrm{T})  -\mu_3(\pmb{\nabla v}_3+\pmb{\nabla v}_3^\mathrm{T})\right]\cdot\pmb{e}_t = \pmb{e}_t \cdot \pmb{\nabla}_t\gamma \\
	\pmb{e}_n\cdot \left[\mu_i(\pmb{\nabla v}_i+\pmb{\nabla v}_i^\mathrm{T}) -\mu_3(\pmb{\nabla v}_3+\pmb{\nabla v}_3^\mathrm{T})  +(p_3-p_i)\pmb{I} \right]\cdot\pmb{e}_n = - \kappa\gamma,
\end{gather}
where $\pmb{e}_n$ and $\pmb{e}_t$ are unit vectors in the normal and tangential direction with respect to the interface of each droplet. The symbol $\pmb{\nabla}_t$ indicate the gradient in the tangential direction. The subscript $i=1,2$ indicated that a quantity is inside the respective droplet. $\pmb{I}$ is the identity matrix and $\kappa$ is the curvature of the interface.

For both droplets the tangential velocities are continuous across the interface,
\begin{equation}
	(\pmb{v}_{3} - \pmb{v}_{i})\cdot\pmb{e_t} = 0.
\end{equation}

Given that the top droplet is considered insoluble in water, no flux takes place across its interface, i.e. $\partial_n c=0$, with $\partial_n$ the partial derivative in the direction normal to the droplet interface. The resulting kinematic condition is
\begin{equation}
	(\pmb{v}_j-\pmb{v}_\mathrm{interf})\cdot \pmb{e}_n = 0.
\end{equation}
where $\pmb{v}_\mathrm{interf}$ is the velocity of the interface and $j=2,3$ (see Figure \ref{fig:SimBC}). On the contrary, the bottom droplet is allowed to dissolve. Therefore its corresponding kinematic condition is
\begin{equation}
	\rho_k(\pmb{v}_k-\pmb{v}_\mathrm{interf})\cdot \pmb{e}_n = j_\mathrm{p} + j_\mathrm{w},
\end{equation}
with $k=1,3$ (see Figure \ref{fig:SimBC}).

The fluxes of 1-pentanol $j_\mathrm{p}$ and water $j_\mathrm{w}$ are also subject to the boundary conditions
\begin{align}
	\rho_3 D_\mathrm{p,w}\partial_n c & = (1-c)j_\mathrm{p} - c j_\mathrm{w}, \\
	\rho_2 D_\mathrm{w,p}\partial_n c_\mathrm{w} &= (1-c_\mathrm{w})j_\mathrm{w} - c_\mathrm{w} j_\mathrm{p}.
\end{align}

The concentrations at the interface were assumed to be the corresponding saturation values on both sides of the interface of the bottom droplet,
\begin{gather}
	c_\mathrm{interf} = c_\mathrm{sat}, \quad \mathrm{bath} \: \mathrm{side},\\
	c_\mathrm{interf,w} = c_\mathrm{sat,w}, \quad \mathrm{drop} \: \mathrm{side},
\end{gather}
with the saturation values obtained from literature \cite{kinoshita1958solubility,IUPAC1984solubility}.

At the bottom and right edge of the domain we considered no-slip boundary conditions and no mass flux.  Finally, an open inflow was assumed for the top edge.

\section{Results}

\subsection{Reference case: dissolution of pentanol droplets \label{subsecCh3:ReferenceCase}}

\begin{figure}
	\centering
	\includegraphics[width=\textwidth]{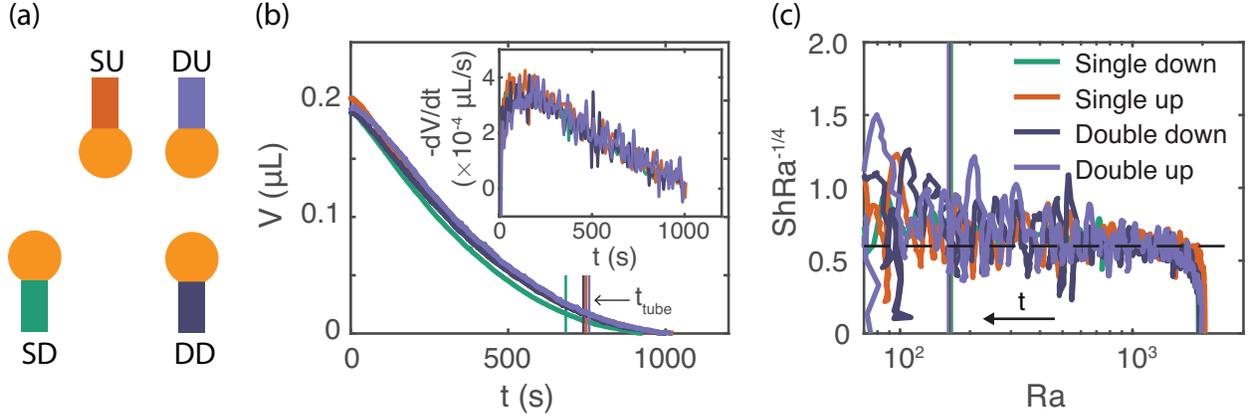}
	\caption{Dissolution of pentanol droplets. (a) Diagrams showing the three configuration used in this figure. A single droplet down (SD), a single droplet up (SU), and double droplets down (DD) and up (DU). (b) Volume versus time of four pentanol droplets in the three configuration described in panel (a). \textit{Inset}: Rate of volume change over time showing a local maximum. (c) Sherwood number compensated with $Ra^{-1/4}$ versus the $Ra$ number. The horizontal dashed black line is a guide to the eye to compare with the data. Data with $Ra < 70$ have been omitted because the influence of the tube increases the noise significantly (the complete data set can be found in the SI). The vertical colored lines in panels (b) and (c) indicate the time $t_\mathrm{tube}$ when the radius of curvature $R_\mathrm{c}$ is equal to the radius of the tubes $R_\mathrm{t}$. }
	\label{fig:PentanolDissolution}
\end{figure}

We first investigated the dissolution of pentanol droplets. We looked at the three different configurations shown in Figure \ref{fig:PentanolDissolution}(a). In the first case we placed an individual droplet with the capillary below it, in the second case the capillary was above the droplet, and in the third case two droplets dissolved simultaneously, one above the other. The initial distance $L_0$ (see Figure \ref{fig:Setup}(a) for a definition) between the two droplets was $L_0\sim2R_{e,0}$, where ${R_{e,0} = (3V_0/(4\pi))^{1/3}}$ is the initial equivalent radius, with $V_0 \sim 0.2 $ \si{\micro\liter} the initial droplet volume.

The droplet was assumed to be a spherical cap, thus to calculate its volume we fitted a circle to the edge of the two-dimensional projection of the droplet. The radius of the circle is the same  as the radius of the spherical cap under the axysimmetric assumption. We found the intersection between the fitted circle and the edge of the capillary to determine the base of the droplet (the procedure is explained in more detail in the SI). The deviation of the projected edge from a circle was small and only noticeable at the beginning. The Bond number $Bo = gR_{e,0}^2\Delta \rho / \gamma$ for the droplets is much smaller than 1 as shown in Table \ref 
{tabCh3:PropertiesOfLiquids}. Those small value are in agreement with at most only a small deformation of the droplet from the spherical cap shape. As dissolution takes place the radius and consequently the Bond number become smaller, making the deviation from the spherical cap shape even smaller.

The experimentally observed droplet volumes for the three configurations are plotted in Figure \ref{fig:PentanolDissolution}(b). There is no significant difference in the time evolution of the volume. While it seems that the single droplet at the bottom dissolves slightly faster than those of the other cases, the deviation is within our experimental error. In fact, from the inset in Figure \ref{fig:PentanolDissolution}(b), it can be seen that the rate of change of the volume over time is essentially the same for the various cases. Therefore, the presence of a second droplet and the capillaries does not seem to considerably affect the dissolution process of the lower droplet.

We calculated the solutal Rayleigh number and the Sherwood number of our system as  \cite{dietrich2016role}
\begin{equation}
	Ra = \frac{g\beta \Delta c R_{e,0}^3}{\nu D}
\end{equation}
and 
\begin{equation}
	Sh = \frac{\langle \dot{m}\rangle_A R}{D\Delta c},
\end{equation}
where $\nu$ is the dynamic viscosity of the bath, $D$ the diffusion coefficient of 1-petanol in water, $c$ the concentration of solute, ${\Delta c = c_\mathrm{sat} - c_\infty = c_\mathrm{sat} }$ the excess concentration at a given position with respect to the bath at infinity (assumed to be pure water), and  $\beta = (\partial \rho/\partial c)/\rho_b$ the solutal expansion coefficient, with $\rho_b$  the density of the bath. $\langle \dot{m}\rangle_A$ is the mass transfer flux, averaged over the surface area of the droplet in contact with the bath.

Previously, it has been shown that the dissolution of sessile droplets with $Ra >12$ is dominated by convection because of the formation of a plume above the droplet \cite{dietrich2016role, chong2020convection}.  As explained in these two references, in the convection dominated regime the Sherwood number should depend on the Rayleigh number as $Sh \sim Ra^{1/4}$. In Figure \ref{fig:PentanolDissolution}(c) we have plotted the $Sh$ number compensated by $Ra^{-1/4}$ versus the $Ra$ number. We can see that after a transient time and while $Ra \gtrsim 100$, the data is approximately flat, indicating that indeed our system is dominated by convection, at least while the droplet is larger than the size of the capillary. The vertical colored solid lines in the Figures  \ref{fig:PentanolDissolution}(b) and (c) indicate the time $t_\mathrm{tube}$ at which the $R_c = R_t$.

We have compared our data to the predictions of diffusion dominated models like the one of Epstein-Plesset \cite{epstein1950stability}, and indeed the dissolution times for those models are much larger than we observe experimentally, as expected from the value of $Ra \gg 12$ \cite{dietrich2016role}.

\color{black}

\begin{figure}
	\centering
	\includegraphics[width=\textwidth]{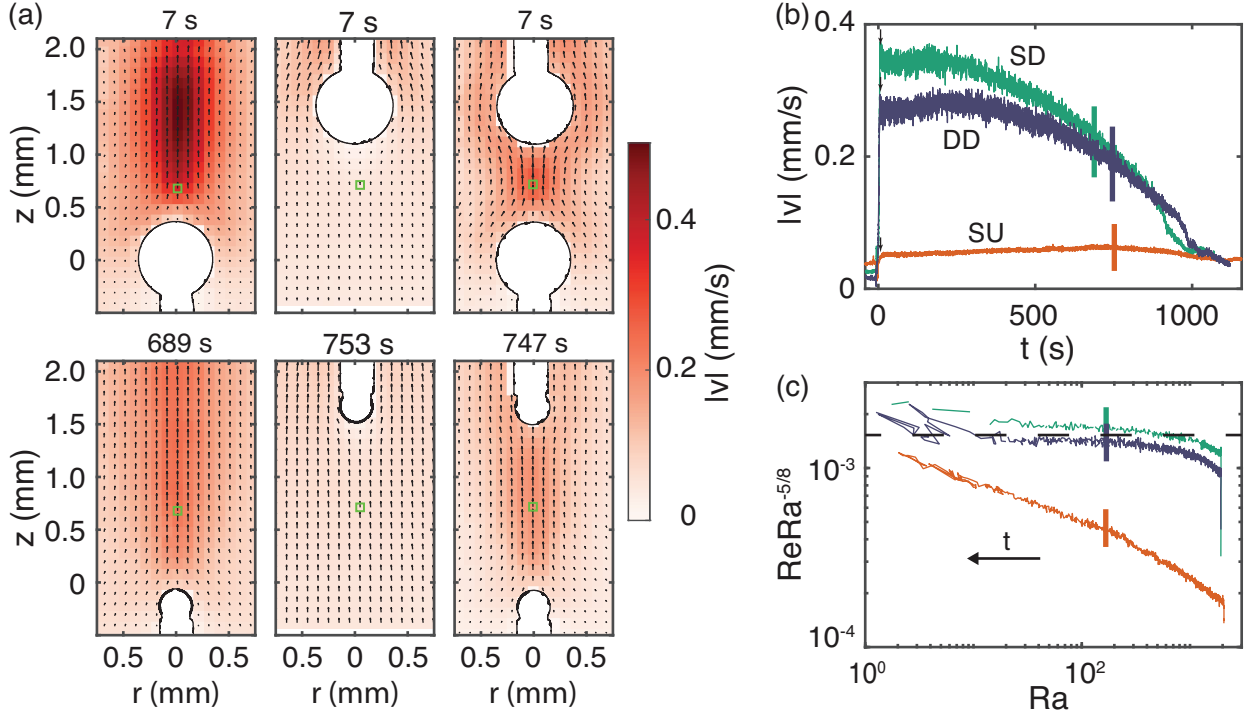}
	\caption{Velocity caused by natural convection. (a) Velocity fields around the three different configutations. Two times are shown: 7 seconds after the droplets were infused (top) and when $t = t_\mathrm{tube}$ (bottom). The velocity fields are averaged over a few frames, equivalent to $\Delta t=1$ \si{\second}. (b) Velocity versus time taken at the axis of symmetry and approximately one initial radius away from the edges of the droplets. The locations where the velocity was obtained are marked with a small green square in the velocity fields of panel (a). The vertical black arrows indicate $t=7$ \si{\second}. (\textit{c}) Reynolds number compensated with $Ra^{-5/8}$ versus the $Ra$ number. The horizontal dashed black line is a guide to the eye to compare with the data. The vertical colored lines in panels (b) and (c) indicate $t = t_\mathrm{tube}$.}
	\label{fig:PentanolVelocity}
\end{figure}

We have also measured the velocity field of the liquid surrounding the droplet. In Figure \ref{fig:PentanolVelocity}(a) we show the velocity fields for the three configurations, with $L_0 \sim 2R_{e,0}$ for the double droplet case. Two times are shown, first a few seconds after the droplets were infused and later when the droplets are the same size as the capillary. In all cases there is the formation of a plume evidenced by the higher velocity above the droplets (not shown for the SU case). For the SU case, the velocity in the region below the droplet is very small, indicating that the mass flux in the plume does not originate from below the droplet but from the sides. This partially explains why the dissolution process is not strongly affected by the presence of the upper droplet. 

We measured the velocity over time at the axis of symmetry and about one initial radius away from the edges of the droplets. Figure \ref{fig:PentanolVelocity}(b) shows the corresponding plot. The velocity first quickly goes up as the plume builds up (first $\sim$ 6 s) and reaches an approximately constant value. However, eventually the velocity has to decrease as the droplet and its Rayleigh number become smaller. According to our data the velocity of the individual lower droplet is at first larger than in the double droplet case, but eventually they equalize. The initial difference must come from the obstacle effect that the upper droplet has, which forces the plume coming from the lower droplet to go around the upper droplet. We have also included the velocity measured one radius below the single upper droplet in Figure \ref{fig:PentanolVelocity}(b). As already mentioned, the velocity is much smaller and is not very different from the existing background flow.

Dietrich et al. \cite{dietrich2016role} also found a scaling law between the Reynolds number of the bath $Re = UR/\nu$ and the $Ra$ number, with $U$ the velocity above the droplet and on the axis of symmetry. In Figure \ref{fig:PentanolVelocity}(c) we have plotted the $Re$ number compensated with $Ra^{-5/8}$ versus the $Ra$ number in a double logarithmic plot. After a transient time, we found the same scaling as Dietrich et al. \cite{dietrich2016role}, at least for an intermediate range of $Ra$ numbers. Here the noise of our data is smaller because we do not have to take a derivative, therefore we can see the trend down to lower values of $Ra$. Obviously, for the SU case the data does not follow the power law since there is no plume below the droplet. 

It is important to notice that the plume is present up to the end of the life of the droplet because there is always 1-pentanol inside the capillary. The plume only disappears once some of the 1-pentanol inside the tube has dissolved. Therefore, we do not observe a strong $Re$ number decrease at low $Ra$ as opposed to Dietrich et al. \cite{dietrich2016role}.

\subsection{Effect of changing the liquid of the upper droplet \label{subsecCh3:DifferentTopLiq}}

\begin{figure}
	\centering
	\includegraphics[width=\textwidth]{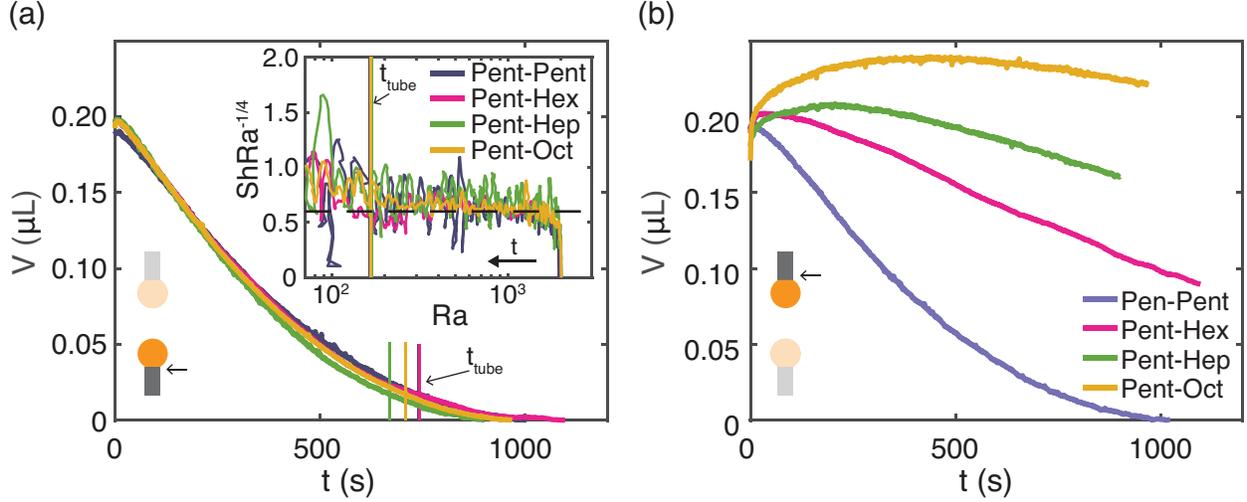}
	\caption{Volume versus time for a 1-pentanol droplet dissolving below a second droplet of 1-pentanol or a different alcohol (1-hexanol, 1-heptanol, or 1-octanol). The initial separation is $L\sim2R_{e,0}$. (a) Volume of the \textit{lower} droplet. The dissolution of the bottom pentanol droplet is hardly affected by changing the alcohol used for the upper droplet. \textit{Inset}: $Sh$ number compensated with $Ra^{-1/4}$ versus the $Ra$ number, calculated with the volume of the lower droplet. For large values of the Ra the curve is approximately flat. The vertical colored lines in both plots indicate the time $t_\mathrm{tube}$ at which the $R_c = R_t$. (b) Volume of the \textit{top} droplet as function of time. When the top droplet is made of a different alcohol, it first increases in volume before it starts to dissolve. The diagrams in both panels (a) and (b) indicate the droplet from which the volume is plotted by the darker color and the arrows.}
	\label{fig:DissolutionDouble}
\end{figure}

With the dissolution of 1-pentanol droplets as a reference, we can now look at the effects of changing the liquid of the top droplet to 1-hexanol, 1-heptanol, or 1-octanol. In Figure \ref{fig:DissolutionDouble}(a) we show that there is no effect in the volume evolution of the bottom droplet, as expected from the results of the previous section. The inset of Figure \ref{fig:DissolutionDouble}(a) shows the non-dimensional numbers based on the volume of the lower droplet. As before, for $Ra \gtrsim100$ the data is approximately flat, while for lower values of $Ra$ the variation in the $Sh$ number are too large to see any clear trend. If we focus our attention to the time when the droplet is larger than the capillary, we can conclude that changing the liquid of the top droplet does not seem to affect the mechanism behind dissolution, namely convection dominated dissolution.

On the contrary, as can be seen from Figure \ref{fig:DissolutionDouble}(b), once the liquid of the top droplet is different from 1-pentanol, there is an initial increase in volume, more clearly seen for 1-heptanol and 1-octanol. This is an indication that some of the 1-pentanol that is transported by the plume dissolves into the upper droplet. To corroborate that this increase in volume is not a result of residual pressure in the tube, we looked at the dissolution of single droplets in water held at the top position. In the single droplet cases there can be indeed a small increase in volume at the beginning which is comparable with the increase observed in Figure \ref{fig:DissolutionDouble}(b) for 1-hexanol. However, in the cases of 1-heptanol and 1-octanol the increase in volume of the single droplet case is considerably smaller than in the double droplet case, indicating that it must be a real effect and not an artifact caused by residual pressure. We show a comparison of the volume time series between the single and double droplet cases in the SI. 

The upper droplet grows for a longer time the longer the carbon chain, as clearly shown by  Figure \ref{fig:DissolutionDouble}(b). The plausible explanation is the decreasing solubility of the alcohol with increasing length of the carbon chain (cf. Table \ref{tabCh3:PropertiesOfLiquids}): The saturation concentration at the interface, and in turn the dissolution rate, will thus be smaller for a longer carbon chain. Since there is no alcohol of the top droplet already dissolved in the bath, dissolution must happen even if 1-pentanol is diffusing into the droplet. The volume change of the droplet will be given by the sum of both fluxes, and since the flux out is stronger for the smaller carbon chain, the volume increases only for a small time. However, as we will show next, there is a Marangoni flow at the top droplet taking place at the start of the dissolution process which could also affect the time over which the droplet grows in size.

\begin{figure}
	\centering
	\includegraphics[width=\textwidth]{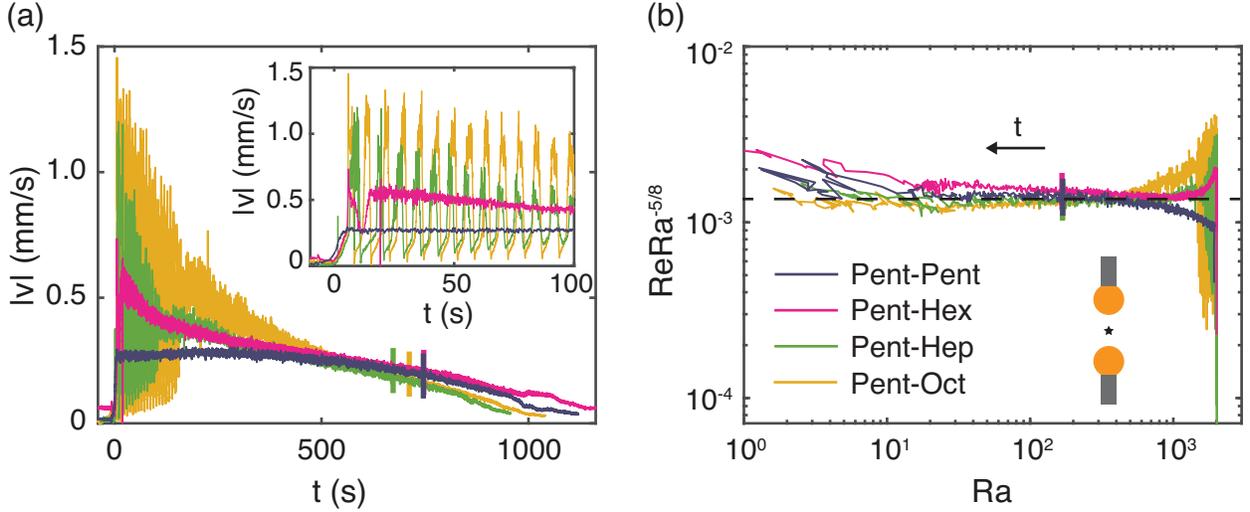}
	\caption{Velocity related data for pairs of droplets dissolving simultaneously with an initial separation of $L_0\sim2R_{e,0}$. (a) Velocity magnitude taken at the middle point between the two droplets. When the liquid of the upper droplet is not 1-pentanol the velocity temporarily oscillates once the plume created by the lower droplet reaches the upper droplet. \textit{Inset}: Zoom of the first 100s. During the oscillations, the velocity reaches up to 1 mm/s and then goes down to close to zero in each cycle. For the first cycles the vertical velocity can even become momentarily negative. (b) Reynolds number compensated with $Ra^{-5/8}$ versus the $Ra$ number. In all cases, there is a region over which the data is approximately flat. The horizontal flat dashed line is only used as a reference. \textit{Inset}: Sketch of the two droplets. The black star indicates the position where the velocity was measured.}
	\label{fig:VelocityDouble}
\end{figure}

In Figure \ref{fig:VelocityDouble}(a) we show the velocity measured one radius above the bottom droplet when the liquid of the top droplet is changed from 1-pentanol to 1-octanol. Surprisingly, the velocity starts to oscillate once the 1-pentanol plume reaches the top droplet. In the inset of Figure \ref{fig:VelocityDouble}(a) we only plotted the first 100 \si{\second} to better visualize the oscillations. Eventually, the oscillation stops and the velocity decreases over time. 

The oscillatory behavior of the velocity is reminiscent of other oscillatory systems like the flows around droplets in a density stratified bath \cite{Li_2019_Bouncing, schwarzenberger2015relaxation}, the up and down movement of a bubble inside a binary bath close to a heated surface \cite{zeng2021periodic}, the periodic emission of plumes of a dissolving droplet \cite{lappa2004higher,lappa2004mixed,lappa2006oscillatory} or the intermittent kicking of a pendant droplet of water surrounded by a binary gas \cite{sutjiadi2008interfacial}. Except for the pendant droplet case, in all the other systems just mentioned, there is an imposed concentration or temperature gradient in the fluid surrounding the droplet, thus gradients of density and surface tension. The competition between these gradients drive the oscillatory flows. In the case of the pendant droplet surrounded by a gas, the gradients in surface tension and density come from the mass transfer from or to the droplet, nevertheless, these gradients are also the cause of the flow inside and outside the droplet.  In our case, given that the 1-pentanol plume creates a gradient in concentration around the upper droplet, we can expect that a similar competition is at place in our experiments.

More precisely, we can imagine that when the plume touches the upper droplet, there is a local dip in the interfacial tension at the bottom of the top droplet. As a consequence, a Marangoni flow develops and pulls liquid from the bottom to the top of the droplet. We indeed see that the velocity suddenly increases in magnitude when the plume reaches the top droplet (see movie 1 in the SI). In the case of the jumping droplets in a stratified flow \cite{Li_2019_Bouncing}, the oil droplet suddenly jumps because of the Marangoni flow caused by the surrounding concentration gradient. Later on, the same Marangoni flow causes the homogenization of the liquid inside the boundary layer around the droplet, stopping the Marangoni flow itself. Afterwards, the droplet moves down due to gravity until diffusion allows the far field stratification to get in contact with the droplet once more, triggering the Marangoni flow again. In a similar way, one could assume that in our system the Marangoni flow transports the pentanol-rich liquid all around the droplet until a homogeneous layer is created around the top droplet. Later, the lower density of the pentanol-rich water as compared to pure water would advect the concentration boundary layer away, bringing the system to the initial state, where the plume could cause a concentration gradient along the interface. However, in the following section we will see, with the help of numerical simulations, that this view is slightly too simplistic and that in fact there is an extra step in which the lower plume detaches from the top droplet, and it is this step which actually causes the oscillatory behavior.

In Figure \ref{fig:VelocityDouble}(b) we show the corresponding compensated $Re$ numbers as function of the $Ra$ number. Again, the data is flat for intermediate values of the $Ra$ number, further suggesting that the 5/8 scaling law becomes applicable only after the droplet dissolution process has been ongoing for some time. In this compensated version of the plot we can also see that for low $Ra$ numbers the data go up, indicating that the capillaries and the liquid within them start to have an effect.

\section{Numerical simulations of the oscillatory regime \label{secCh3:Oscillations}}

\begin{figure}
\centering
\includegraphics[width=\textwidth]{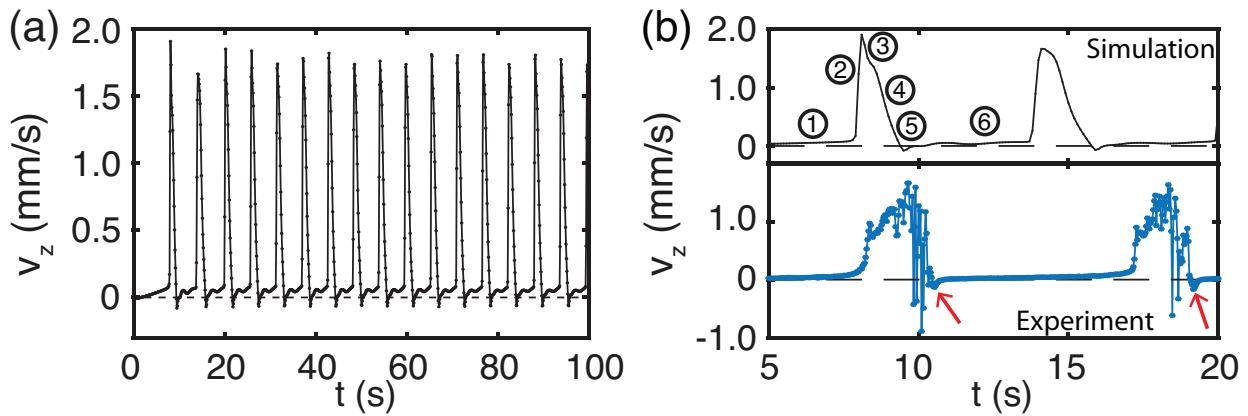}
\caption{(a) Vertical velocity at the middle point between the two simulated droplets for 100 \si{\second}. The initial separation between the edges of the droplets is $L_0=R_{e,0}$. In the simulation the oscillations are present for the whole life of the lower droplet. (b) Comparison of the central velocity between simulations (upper) and experiments (lower). The qualitative behavior is satisfactorily recovered in the numerical simulation and also the order of magnitude of the velocity is in good agreement. The period of one cycle in experiments is slightly longer than in the simulations, which we attribute to the simplifications taken in the numerical model (see text). The 6 times marked in the upper plot of panel (b) indicate the corresponding times of the snapshots shown in Figure \ref{fig:OneCycleDescription}.}
\label{fig:CentralVel}
\end{figure}

We simulated the case of a 1-pentanol down and a 1-octanol droplet up with an initial separation of $L_0=R_{e,0}$. In Figure \ref{fig:CentralVel}(a) we have plotted the vertical velocity obtained at the middle point between the droplets. The oscillatory behavior of the velocity is satisfactorily recovered. In Figure \ref{fig:CentralVel}(b) we show a zoom to the first two cycles and compare them to the first two cycles obtained experimentally ($L_0=R_{e,0}$). It can be seen that the maximum velocity in both simulations and experiments are of the same order of magnitude, with the velocity in simulations being slightly larger. As for the period of oscillation, while it is longer in experiments, both have the same order of magnitude. The fair qualitative and to some degree even quantitative agreement lets us conclude that the diffusion into and from the upper droplet, which is turned off in the simulation, is not a key element for the appearance of the oscillatory behavior.

\begin{figure}
	\centering
	\includegraphics[width=\textwidth]{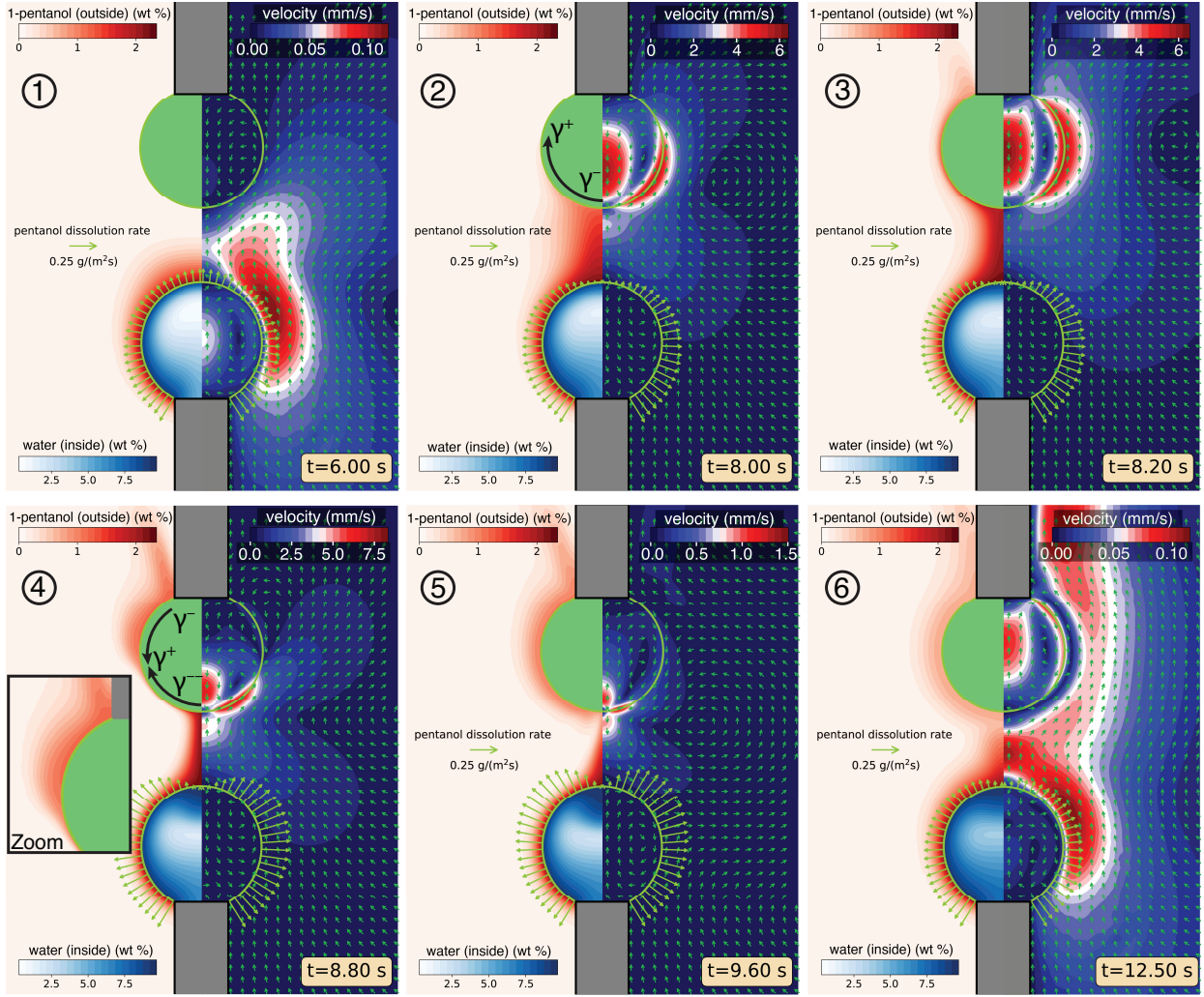}
	\caption{Snapshots of the numerical results at 6 different moments during one oscillation cycle. At \protect \circled{1}, the plume is building up and moves towards the top droplet. At \protect \circled{2}, the plume has reached the top droplet, creating a gradient in interfacial tension and triggering an upwards Marangoni flow. At \protect \circled{3}, the Marangoni flow has brought 1-pentanol rich liquid all around the droplet in a relatively thin concentration boundary layer. At \protect \circled{4}, pentanol rich liquid has accumulated close to the contact line between the top droplet and the top capillary, causing a secondary Marangoni flow. A zoom to the upper droplet shows the accumulation. At \protect \circled{5}, the secondary Marangoni flow has created a thicker concentration boundary layer all around the droplet. The velocity at the symmetry axis and in between the droplets points down, causing the plume to “detach" from the upper droplet. At \protect \circled{6}, buoyancy competes with the secondary Marangoni flow. Eventually the plume is able to get in contact with the upper droplet again and the process restarts.}
	\label{fig:OneCycleDescription}
\end{figure}

From the simulation, we can further verify the mechanism behind the oscillations proposed in the previous section. We have selected 6 different moments along the first oscillation cycle from the simulation, as marked in Figure \ref{fig:CentralVel}(b) . The corresponding snapshots are shown in Figure \ref{fig:OneCycleDescription}. 

As we already concluded from experiments, initially (at time \circled{1}) the dissolution of pentanol causes an accumulation of pentanol rich liquid around the bottom droplet, that results in the creation of a convective plume. In the snapshot at time \circled{1} the plume is just starting to move up. A couple of seconds later, the plume gets in contact with the top droplet as shown in snapshot \circled{2}. There is a gradient in concentration along the interface with the higher concentration at the bottom of the droplet. Therefore the interfacial tension increases from the bottom towards the top of the droplet and a Marangoni flow appears in the same direction. Just a fraction of a second later, at time \circled{3}, the Marangoni flow extends along the whole droplet and the pentanol-rich liquid has been transported to the top of the droplet.

From our hypothesis in the previous section, the Marangoni flow could help to homogenize the pentanol-rich liquid around the top droplet, therefore stopping itself. However, from the snapshot at time \circled{4}, we can see that before homogenization takes place, a contrary gradient in 1-pentanol is created at the top part of the droplet (see the zoomed region). Therefore, an adversary Marangoni flow is created, evidenced by the velocity field inside the droplet. The opposing flow brings back down the pentanol-rich liquid and increases the thickness of the concentration boundary layer around the top droplet. Eventually, the velocity inside the droplet completely reverses and the secondary Marangoni flow takes over (time \circled{5}). The downward flow pushes the plume away from the top droplet, letting the secondary Marangoni flow dominate. Notice that by bringing down pentanol-rich fluid and increasing the thickness of the boundary layer, the result is similar to the homogenization caused by diffusion in the bouncing droplet case \cite{Li_2019_Bouncing}.

Despite the downward flow caused by the secondary Marangoni flow, the pentanol-rich fluid is less dense than water, thus buoyancy eventually takes over and, as shown in time \circled{6}, the flow starts to move upward once more. This causes that outside the top droplet and very close to its interface the velocity points down, but after a small distance it points up (look at the equator of the droplet at time \circled{6} in Figure \ref{fig:OneCycleDescription}). Eventually, the plume is able to reach the top droplet again and the whole process repeats.

When we look closely at our experiments, we can see that indeed there is a momentary downward flow at each cycle (at least for a few first cycles) in accordance with what we see in the simulations. We point at the negative flow in Figure \ref{fig:CentralVel}(b) with red arrows, which can also be seen in the movie in the SI. 

\section{Discussion \label{secCh3:Discussion}}
A possible explanation for the accumulation of 1-pentanol at the top of the droplet is the no-slip boundary condition on the top capillary. As we can see from the simulations, at time \circled{3}, the pentanol-rich solution does not go up parallel to the tube wall. Instead, it is advected at an oblique angle due to the velocity boundary layer created at the wall. The concentration field takes the shape of a horn. The gradient in pentanol mass fraction in the normal direction to the interface, $\partial_n c$, and hence diffusion, is smaller at the position of the horn than in other parts on the interface. This difference in diffusive flux can cause that the concentration at the location of the horn becomes higher than at other points on the interface. Once this happens, the contrary Marangoni flow appears and the thicker concentration boundary layer starts to propagate down the interface. 

One can then wonder whether the oscillatory behavior would take place if the capillary was removed, because the no-slip boundary condition would not be present. We can still expect that a stagnation point would form on the top of the droplet and advection could still create a ``horn" but this time on the top of the droplet. Therefore, we can speculate that even without the capillary the oscillatory behavior could appear. Moreover, if the Reynolds number were to be larger, and a pair of standing eddies were to appear behind the droplet, they would serve as a way to accumulate pentanol-rich solution on top of the droplet and trigger the secondary Marangoni flow. However, for the moment this is only a speculation.

As opposed to the experiments, in the simulation, the oscillations lasted for the whole life of the lower droplet. This suggests that some of the simplifications taken in the simulation play a role in determining the time during which the oscillatory behavior is active. First of all, in the simulation we did not consider the dissolution of 1-pentanol into the 1-octanol droplet, which is clearly happening in experiments. At the same time, 1-octanol can also dissolve into the bath. Both mass transfer processes can affect the concentration at the interface and thus the Marangoni flow. Moreover, it is known that mass transfer of a solute that affects the interfacial tension can destabilize an interface \cite{SternlingScriven1960AIChE, Kovalchuk2006marangoni, Kollner_2013_multiscale, Schwarzenberger_2014}. Additionally, the liquids in the simulation are pure, while in experiments there can always be some contamination that could stop the flow. Some other possible effects that are not taken into account in the simulations are temperature variations, the presence of 1-pentanol inside the capillaries, and pressure perturbations coming from the capillaries and the syringes.

We have performed a few experiments and numerical simulations with different initial distance between the droplets, namely $L_0 = R_{e,0}$, $L_0=2R_{e,0}$, $L=4R_{e,0}$. The general trend is that for a larger spacing, the number of oscillations decreases (see the SI). Possibly this is a consequence of a weaker concentration gradient in the plume, both in the axial and radial directions, because the plume travels for a longer distance. A smaller gradient results in a smaller Marangoni flow. Surprisingly, in a simulation with $L_0= 2R_{e,0}$ the oscillations were reduced to only 3 cycles, while in experiments the average number of oscillations was 22 at the same spacing. As already discussed, this indicates that some of the simplifications taken in the simulations affect the number of oscillations, apart from the initial separation distance $L_0$.

When the liquid is changed to a shorter carbon chain alcohol, the derivative of the surface tension with respect to the concentration of 1-pentanol ($\partial \gamma/\partial c$) is reduced (cf. Table \ref{tabCh3:PropertiesOfLiquids}). As we saw in the previous section, oscillations did not happen when the top droplet was made of 1-pentanol, or they were very weak and for only a few cycles for 1-hexanol (see Figure \ref{fig:VelocityDouble}(a)), indicating that indeed a strong enough interfacial tension gradient is needed to produce oscillations.

Additionally, changing the geometry at the contact line also seems to have an effect on the total number of cycles, as we discuss in the SI. Clearly, there are many parameters to be considered, but making an exhaustive exploration of the parameter space is beyond the scope of the present work.

\section{Conclusions and outlook \label{secCh3:Conclusion}}

We have studied the dissolution of droplets of alcohols with different lenghts of carbon chains inside a water bath, mainly focusing on a system of two vertically aligned droplets dissolving simultaneously. As a reference case we looked at the dissolution of an individual droplet of 1-pentanol and of two vertically aligned 1-pentanol droplets. We built up from these base cases and changed the liquid of the top droplet to alcohols with longer carbon chains. In all the cases, considering the dissolution of the lower droplet and the velocity of the plume, we recovered the scalings $Sh \sim Ra^{1/4}$ and $Re \sim Ra^{5/8}$. The scalings appear after an initial transient period and for large enough Rayleigh numbers, in particular, as long as the droplets are larger than the capillary used to hold them. This is in agreement with the results obtained previously for the dissolution of sessile droplets  \cite{dietrich2016role,chong2020convection}.

For two vertically aligned droplets, when the liquid of the top droplet was different from 1-pentanol, the interaction of the upper droplet with the 1-pentanol plume coming from below triggered an oscillatory Marangoni flow. During one cycle, the Marangoni flow pulls the liquid upwards, in the same direction as natural convection. However, the presence of the top capillary and the no-slip condition at its walls result in an inversion of the concentration gradient along the interface of the droplet. Consequently, an adverse secondary Marangoni flow is triggered, which competes with the primary Marangoni flow and with natural convection. The secondary flow hinders the flow of the 1-pentanol plume and detaches the top droplet from the plume. Eventually buoyancy takes over and brings the plume back into contact with the upper droplet again. 

By comparing our experimental and numerical results we could conclude that the diffusion of 1-pentanol into the top droplet and the dissolution of the top droplet into the bath do not play a prominent role in triggering the oscillatory behaviour. However, mass transfer can affect the total duration of the oscillatory regime and causes a temporary increment in the volume of the top droplet. While the parameter space is large, we were able to recognize that for a larger spacing between the droplets the oscillatory behavior lasts less. A reduction on the value of the derivative of interfacial tension with respect to the concentration can also eliminate the oscillatory behavior. In future works, it would be interesting to further explore the parameter space by including more values of the initial distance $L_0$ between the droplets and of the initial radii. It would also be instructive to change the density difference between the plume and the bath, and the viscosity of the liquids. In addition, it would be very interesting to see whether the oscillations persist if for example the capillary holds the upper droplet from the side and not from the top, or if in simulations the capillary is completely removed.

\section{Acknowledgments}
\begin{acknowledgments}
The authors thank Alvaro Marin and many other members of the Physics of Fluids group for valuable discussions about the experimental setup and in particular Yanshen Li for a very insightful discussion on the mechanism behind the mixing caused by the Marangoni flow.
\end{acknowledgments} 

\section{Funding}
This work was supported by the EU (ERC-Advanced Grant Project DDD No. 740479, and ERC-Proof-of-Concept Grant Project No. 862032). X.H.Z. acknowledges support from the Natural Science and Engineering Council of Canada (NSERC) and from the Canada Research Chairs program.

\section{Conflict of interest}
There is no conflict of interest for this work.

% Create the reference section using BibTeX:
\bibliography{Bib2Droplets_6p1.bib}

\end{document}